\documentclass[aps,pre,superscriptaddress,showfacs,twocolumn,showpacs]{revtex4}
\usepackage{amsmath}
\usepackage{graphicx}
\usepackage{amssymb}
\usepackage{bbm, color}

\newcommand*{\cl}[1]{{\mathcal{#1}}}

\newcommand*{\tn}[1]{{\textnormal{#1}}}
\newcommand{\ket}[1]{\left|#1\right>}

\begin{document}

\title{
Behavior of three modes of decay channels and their self-energies \\
of elliptic dielectric microcavity
}

\author{Kyu-Won Park}
\affiliation{Department of Physics, Sogang University, Seoul 04107, Korea}
\author{Jaewan Kim}
\affiliation{School of Computational Sciences, Korea Institute for Advanced Study, Seoul 02455, Korea}
\author{Kabgyun Jeong}
\email{kgjeong6@snu.ac.kr}
\affiliation{Center for Macroscopic Quantum Control, Department of Physics and Astronomy, Seoul National University, Seoul 08826, Korea}
\affiliation{School of Computational Sciences, Korea Institute for Advanced Study, Seoul 02455, Korea}

\date{\today}

\begin{abstract}
The Lamb shift (self-energy) of an elliptic dielectric microcavity is studied. We show that the size of the Lamb shift, which is a small energy shift due to the system-environment coupling in the quantum regime, is dependent on the geometry of the boundary conditions. It shows a global transition depending on the eccentricity of the ellipsis. These transitions can be classified into three types of decay channels known as whispering-gallery modes, stable-bouncing-ball modes, and unstable-bouncing-ball modes. These modes are manifested through the Poincar\'{e} surface of section with the Husimi distribution function in classical phase space. It is found that the similarity (measured in Bhattacharyya distance) between the Husimi distributions below critical lines of two different modes increases as the difference of their self-energies decreases when the quality factors of the modes are on the same order of magnitude.
\end{abstract}
\maketitle

%%%%
\section{Introduction} \label{intro}
Understanding an open effect on a quantum system, i.e., coupling the system to its environment, is very important in a dielectric microcavity. For examples, Fresnel filtering~\cite{SHF11,RTSCS02}, the Goos-H\"{a}nchen shift~\cite{SH06,UWH08}, the quasiscar~\cite{LRR+04}, and the exceptional point~\cite{LYM+09,H04,MR08} are only possible in open quantum systems. As well, most typical (real) systems are open systems interacting with their environment, and they are in contrast to the perfect closed (or so-called billiard) system~\cite{KKS99,S99,WWD97}, which is established by the infinite-potential well.

A good way to investigate the open nature of a quantum system is to look into the phenomenon known as the Lamb shift, since it formally deals with differences between closed and open systems. The Lamb shift is a tiny energy transition of a quantum system originating from the system-environment coupling or caused by the vacuum fluctuations~\cite{LR47,SS10}. This Lamb shift is initially observed for a hydrogen atom~\cite{LR47}, but these effects have been generalized to cavity QED~\cite{WKG04}, photonic crystals~\cite{N10}, and also a circular dielectric microcavity~\cite{PKJ16}. Especially, the non-Hermitian Hamiltonian~\cite{R09,D00,CK09,WKH08} via Feshbach projection-operator formalism was introduced to define the Lamb shift in a dielectric microcavity; the difference of energy eigenvalues between the effective non-Hermitian Hamiltonian and the Hermitian Hamiltonian for a closed system is first defined as a Lamb shift in Ref.~\cite{PKJ16}.

Recently, it is known that there are two kinds of Lamb shift~\cite{SS10,RWSS12,RSSCR10}. The first one is \emph{self-energy}, that is, the Lamb shift in atomic physics,
and the other is a \emph{collective} Lamb shift. The considerable difference of these two Lamb shifts will be clear soon, when we explicitly write down the matrix elements of an effective non-Hermitian Hamiltonian.
Here, the diagonal terms are corresponding to self-energy itself and the off-diagonals are directly corresponding to the collective Lamb shift~\cite{R13}.
In this paper, we consider a dependence on the geometrical boundary conditions of the self-energy by global transitions,
and compare between the relative difference of self-energy and similarity of decay channels by Bhattacharyya distance.

This paper is organized as follows. We introduce a Poincar\'{e} surface of section (PSOS) for boundary conditions in an elliptic billiard in Sec.~\ref{PSoS}. In Sec.~\ref{Lamb}, we briefly review the non-Hermitian Hamiltonian and two kinds of Lamb shift. The dependence on geometrical boundary conditions of self-energy and global transitions is presented in Sec.~\ref{decay}. Our main result of crossings of self-energies is discussed in Sec.~\ref{Husimi}. Finally, we conclude the paper in Sec.~\ref{conclusion}.

%%%%
\section{Canonical coordinate and Poincar\'{e} surface of section} \label{PSoS}
\begin{figure}
\centering
\includegraphics[width=9.0cm]{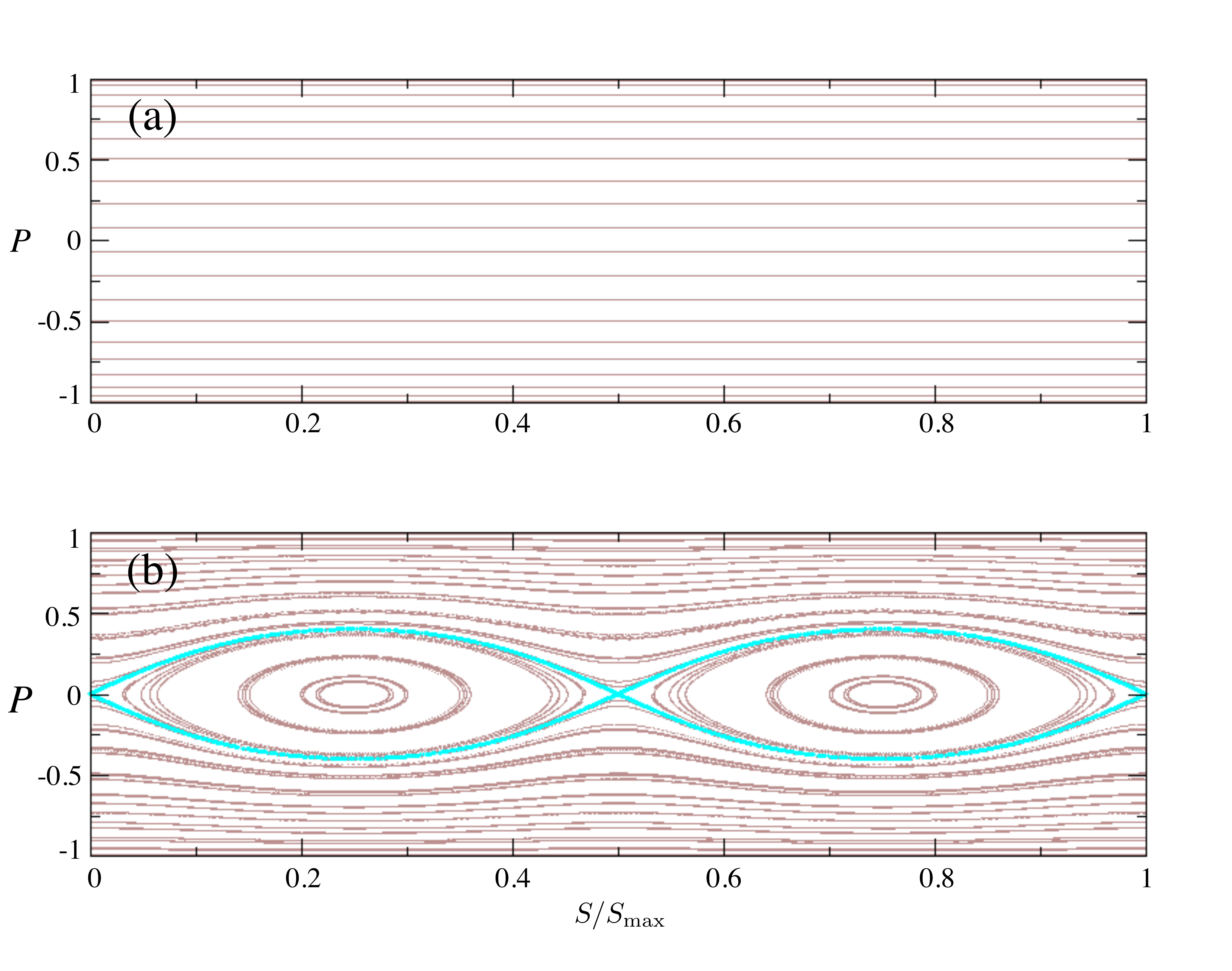}
\caption {(Color online) Poincar\'{e} surface of section. (a) Poincar\'{e} surface of section for a circle. Because of the angular momentum conservation, the tangent momentum $P:=\sin\chi$ is a straight line depending on arc length $S$. (b) Poincar\'{e} surface of section for an ellipse. There are primarily two types of region. One is whispering-gallery motions, and another one is bouncing-ball motions: These two regions are divided by the separatrix whose color is cyan. The bouncing-ball motions also are classified into the stable-island motion and unstable-saddle motion.}
\label{Figure-1}
\end{figure}

The Poincar\'{e} surface of section (PSOS)~\cite{T89} is very useful tool to analyze a ray dynamics of microcavities. For given boundary conditions of a convex billiard system, we can efficiently record not only the sequences of bouncing points $S$ along the boundary wall but also the successive values of $\sin\chi$, where $\chi$ is the incident angle with respect to the normal to the boundary wall. Then we obtain pairs of sequences of points $(S,\sin\chi)$ corresponding to PSOS (see Fig.~\ref{Figure-1}). This plot represents a section through the classical phase space, not the entire phase space itself. So we can analyze the ray dynamics easily, since we deal with it under a one-lower dimension.

Furthermore, $\sin\chi$ is proportional to the tangential component of the momentum at each collision with the boundary, and this momentum component is the canonical conjugate momentum with respect to the boundary length $S$~\cite{N97}.

We here plot the PSOS for a circle in Fig.~\ref{Figure-1}(a). In the case of a circle, since the incident angles and reflection angles are the same to have angular momentum conservation, the tangent momentum $P=\sin\chi$ is a straight line depending on arc length $S$. There is a PSOS for an ellipse at eccentricity $e=0.4$ in Fig.~\ref{Figure-1}(b), which is mainly two types of regions. One corresponds to a whispering-gallery motion and the other is a bouncing-ball motion. These two regions are divided by separatrix whose color is cyan. The bouncing-ball motions also can be divided into stable-island motion and unstable-saddle motion.

%%%%
\section{Non-Hermitian Hamiltonian and two kinds of Lamb shift} \label{Lamb}
Now, we consider a time-independent Schr\"{o}dinger equation with a whole space composed of two subsystems as follows:
\begin{align} \label{schrodinger}
H_T\ket{E}_{AB}=E\ket{E}_{AB},
\end{align}
where $H_T$ is total (Hermitian) Hamiltonian with real energy eigenvalue $E$, and $\ket{E}_{AB}$ represents the corresponding eigenvector of $E$ on a given total system $AB$. For convenience, the subspaces $A$ and $B $ denote a quantum system and an environment (or bath), respectively.

The first subspace is a discrete state of quantum system $A$ and the second one is a continuous scattering state of environment $B$ such that  projection operators, $\pi_{A}$ and $\pi_{B}$, satisfy $\pi_{A}+\pi_{B}={\bf I}_{T}$ and $\pi_{A}\pi_{B}=\pi_{B}\pi_{A}=0$.
Here, $\pi_{A}$ is a projection onto a quantum system whereas $\pi_{B}$ is a projection onto the environment. The operator ${\bf I}_{T}$ is an identity operator defined on total space $AB$. With these projection operators, we can define block matrices such as $h_{A}=\pi_{A}H_T\pi_{A}$, $h_{B}=\pi_{B}H_T\pi_{B}$, $V:=V_{AB}=\pi_{A}H_T\pi_{B}$, and $V^\dag:=V_{BA}=\pi_{B}H_T\pi_{A}$.

The total Hamiltonian in Eq.~(\ref{schrodinger}) can be represented by a block matrix form
\begin{equation} \label{totalH}
H_T=h_{A}+h_{B}+V_{AB}+V_{BA},
\end{equation}
where $h_{A}$ and $h_{B}$ are the Hamiltonian of the system and environment, and $V_{AB}$ and $V_{BA}$ are interaction Hamiltonians between the system and the environment, respectively. The total wave function is also given by
\begin{equation}
\ket{E}_{AB}=\pi_{A}\ket{E}_{AB}+\pi_{B}\ket{E}_{AB}:=\ket{E}_A+\ket{E}_B.
\end{equation}
It is important to note that $\pi_{A}\ket{E}_{AB}=\ket{E}_{A}$ and $\pi_{B}\ket{E}_{AB}=\ket{E}_{B}$. By using these relations, the Hamiltonian eigenvalue problem of Eq.~(\ref{schrodinger}) can be rewritten~\cite{R09,D00,CK09,WKH08,M11} in the form of
\begin{align}
(h_{B}-E)\ket{E}_{B}&=-V_{BA}\ket{E}_{A}\;\;\tn{and} \nonumber\\
(h_{A}-E)\ket{E}_{A}&=-V_{AB}\ket{E}_{B}.
\end{align}
Also note that the states restricted on $A$ and $B$ after the projections are given by~\cite{R09}
\begin{align}
\ket{E}_{B}&=\ket{\cl{E}}+G^{\rightarrow}_{B}V_{BA}\ket{E}_{A}\;\;\tn{and} \nonumber\\
\ket{E}_{A}&=\frac{V_{AB}}{E-H_\tn{eff}}\ket{\cl{E}}, \label{ev_A}
\end{align}
where $\ket{\cl{E}}$ is an eigenvector of $h_{B}$ and $G^{\rightarrow}_{B}$ is an outgoing Green's function in the subspace $B$. The Eq.~(\ref{ev_A}) means that the wave function localized in subsystem $A$ can be obtained through the incoming wave $\ket{\cl{E}}$ penetrating into the subsystem $A$ via coupling term $V_{AB}$ and propagating by effective Green's function $(E-H_\tn{eff})^{-1}$.

Then, by using the total Hamiltonian in Eq.~(\ref{totalH}), we can define the effective non-Hermitian Hamiltonian as
\begin{align}
H_\tn{eff}&=h_{A}+V_{AB}G_B^{\rightarrow}V_{BA} \\
&=h_{A}-\frac{1}{2}iVV^\dagger+P_\tn{v}\int \frac{VV^\dagger}{{E}-\tilde{E}}d\tilde{E}, \label{NHeff}
\end{align}
where $P_\tn{v}$ means the (Cauchy) principal value depending on each decay channel.

\begin{figure}
\centering
\includegraphics[width=9.0cm]{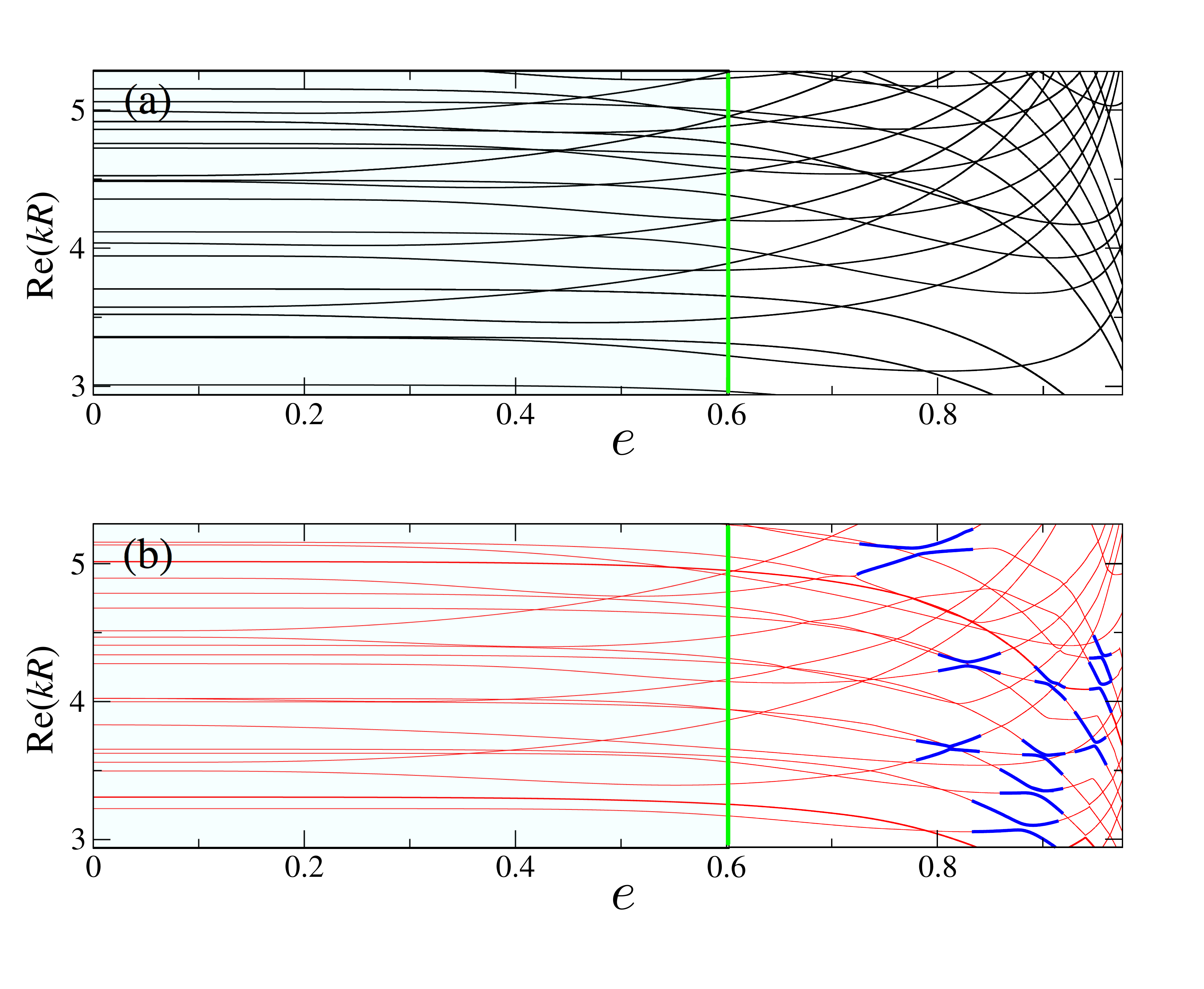}
\caption {(Color online) The plot of $\tn{Re}(kR)$ versus the eccentricity $e$ in an ellipse. (a) The real value of $kR$ in eigenvalue trajectories of the elliptic billiard depending on the eccentricity $e$ from $e=0$ (circle) to $e=0.99$ (ellipse). There are not any avoided crossings in this panel, since the elliptic billiard system is one of the integrable systems. (b) The real values of $kR$ in eigenvalue trajectories of the elliptical microcavity for $n=3.3$ for the refractive index of the InGaAsP semiconductor microcavity depending on the eccentricity $e$ from $e=0$ to $e=0.99$. In contrast to the elliptic billiard system, there are several avoided crossings marked by the thick (blue) lines beyond $e=0.6$ in the open elliptic system. We set the vertical line in at e$=0.6$ to separate the regions with avoided crossings and without avoided crossings.}
\label{Figure-2}
\end{figure}

This non-Hermitian Hamiltonian has generally complex eigenvalues under the Feshbach projection-operator formalism (see Ref.~\cite{PKJ16} and the references therein),
\begin{align}
H_\tn{eff}\ket{\psi_{k}}=\varepsilon_{k}\ket{\psi_{k}}
\end{align}
and its eigenvalues of $H_\tn{eff}$ are given by (for each $k$)
\begin{equation}
\varepsilon_k=E_k-\frac{i}{2}\gamma_k.
\end{equation}
Here, $\varepsilon_{k}$ is an eigenvalue relating to $E_k$ and $\gamma_k$, which represents the energy and decay width of the $k$th eigenvector~\cite{R09,D00,CK09,WKH08}, respectively. The quality factor $Q$ is defined by $\frac{E_k}{2\gamma_k}$.

The Lamb shift, which is a small energy shift mentioned above, can be also obtained by the effective non-Hermitian Hamiltonian in Eq.~(\ref{NHeff})~\cite{PKJ16,R13}. That is,
\begin{align}
%Re\{H_{eff}\}=H_{ss}+\textit{P}\int d\cl{E}' \frac{VV^\dagger}{\cl{E}-\cl{E}'} \\
\Delta H_\tn{Lamb}:=\tn{Re}(H_\tn{eff})-h_{A}=P_\tn{v}\int \frac{VV^\dagger}{E-\tilde{E}}d\tilde{E}.
\end{align}
If we consider the case of a two-level system, then the Lamb shift can be obtained from a $2\times2$ matrix as a toy model:
\begin{equation}
%P_\tn{v}\int \frac{VV^\dagger}{E-\tilde{E}}d\tilde{E}
\Delta H_\tn{Lamb}
=\begin{pmatrix}
\omega_{11} & \omega_{12} \\
\omega_{21} & \omega_{22}
\end{pmatrix}.
\end{equation}
In general, the so-called \emph{self-energy} $S_{e}$~\cite{SS10} due to the diagonal component $\omega_{jj}$ is known as the ``Lamb shift" in atomic physics. On the other hand, the off-diagonal terms $\omega_{jk}\;(\forall j,k=\{1,2\})$ are known as the ``collective Lamb shift" by Rotter~\cite{R13}. Therefore, the collective Lamb shift leads to avoided resonance crossings (ARCs)~\cite{RLK09,J06,SGWC13}, obviously. So we can easily discriminate these two kinds of Lamb shift as shown in Fig.~\ref{Figure-2}. There are two kinds of real values of $kR$ in the eigenvalue trajectories depending on the eccentricity $e$ from $e=0$ (circle) to $e=0.99$ (ellipse). The black solid lines in Fig.~\ref{Figure-2}(a) are the eigenvalue trajectories of $\tn{Re}(kR)$ of the elliptic billiard, and the red solid lines in Fig.~\ref{Figure-2}(b) are those of the elliptic microcavity interacting with its environment (open quantum system). According to the random matrix theory~\cite{H01}, the integrable system, which has $N$ quantum numbers in $N$ degrees of freedom, is followed by a Poisson distribution resulting without avoided crossing. This fact is well confirmed by Fig.~\ref{Figure-2}(a). There are always level crossings even at large eccentricity, i.e., $e=0.99$.

On the contrary, we can check that there are several avoided crossings marked by thick (blue) lines in the elliptic microcavity beyond $e=0.6$. These resonance repulsions must originate from the off-diagonal term $\omega_{jk}$, i.e., the collective Lamb shift. As a result, we can distinguish between the self-energy region and collective Lamb shift region by checking the existence of avoided crossings. In order to consider only the self-energy region, we set a (green) vertical line at $e=0.6$. We remark that a study on the collective Lamb shift will be published soon.

%%%%
\section{Global transition of self-energy and decay channel} \label{decay}
\begin{figure}
\centering
\includegraphics[width=9.0cm]{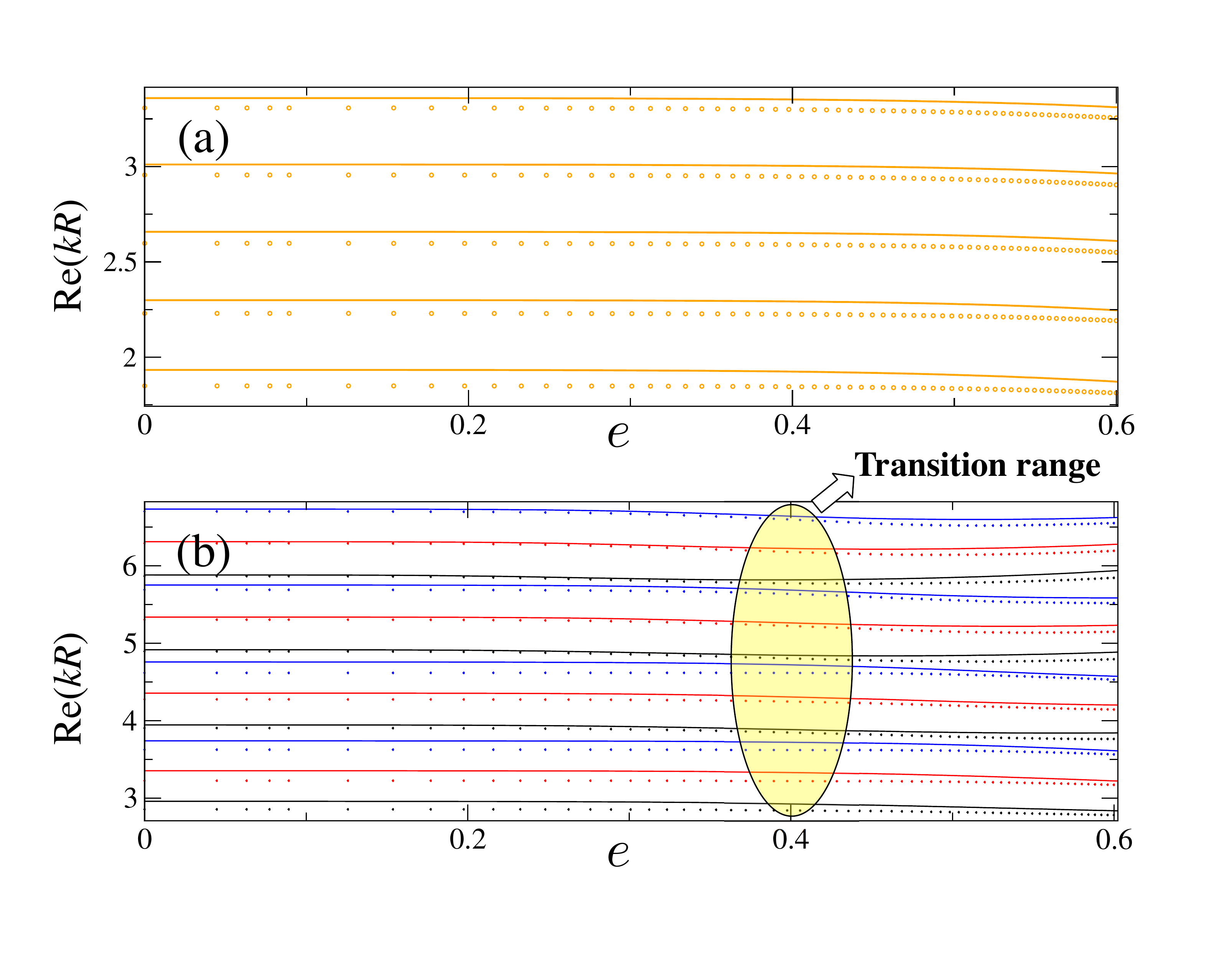}
\caption{(Color online) (a) The real $kR$ in eigenvalue trajectories of the single-layer whispering-gallery mode of $\ell=1$ and $m=(3,4,5,6,7)$ depending on the eccentricity $e$ in both the elliptic billiard and microcavity. The solid lines are eigenvalues of the elliptic billiard, and dotted lines are those of the elliptic microcavity. The difference between the two eigenvalues is almost unchanged even at $e=0.6$. (b) The real $kR$ in eigenvalue trajectories of $(m=3,\ell=2,3,4,5,6)$, $(m=4,\ell=2,3,4,5,6)$, and $(m=5,\ell=2,3,4,5,6)$, respectively, depending on the eccentricity $e$ in both elliptic billiard and microcavity. In contrast to the single-layer whispering-gallery mode, the difference between the two eigenvalues is drastically changed at $e=0.4$.}
\label{Figure-3}
\end{figure}

The Lamb shift was first observed in the hydrogen atom~\cite{LR47}, and then it was studied in cavity QED~\cite{N10} and photonic crystals~\cite{WKG04}, and so on. As mentioned before, the difference of energy eigenvalues between the effective non-Hermitian Hamiltonian in a dielectric microcavity and the Hermitian Hamiltonian for a closed system was first reported as a Lamb shift in Ref.~\cite{PKJ16}.  In this section, we investigate the geometrical boundary dependence of the Lamb shift (self-energy) through global transitions depending on $e$. There are real values $kR$ in eigenvalue trajectories of single-layer whispering-gallery modes (WGMs) of $\ell=1$ and $m=(3,4,5,6,7)$ for orange lines depending on the eccentricity $e$ in both elliptic billiard and microcavity, respectively, in Fig.~\ref{Figure-3}(a). The solid lines are eigenvalue trajectories of an elliptic billiard whereas the dotted lines are those of the elliptic microcavity. We can easily check that the relative differences of the two eigenvalue trajectories are almost unchanged even at the eccentricity $e=0.6$. In Fig.~\ref{Figure-3}(b), there are the real $kR$ in eigenvalue trajectories of $(m=3,\ell=2,3,4,5,6)$, $(m=4,\ell=2,3,4,5,6)$, and $(m=5,\ell=2,3,4,5,6)$, respectively, and depending on the eccentricity $e$ in both elliptic billiard and microcavity. The black lines are for $m=3$, the reds ones are for $m=4$, and the blue ones are for $m=5$. In contrast to the single-layer WGMs, the relative differences between two kinds of eigenvalue trajectories are drastically changed near $e=0.4$. As we defined self-energy in Sec.~\ref{Lamb}, these relative differences of two kinds of eigenvalue trajectories are the self-energy ($S_{e}$).

These patterns of the self-energy transition depending on the eccentricity $e$ are clearly shown in Fig.~\ref{Figure-4} below. We here notice that there are primarily three types of the self-energy transitions represented by increasing group ($\vartriangle$), decreasing group ($\triangledown$), and unchanging group ($\triangleright$). In the case of increasing and decreasing groups, the figures also show that the transition rate of the self-energy is very small until $e=0.3$, but rapidly increasing beyond the eccentricity $e=0.3$. Furthermore, we can also observe that the self-energy of the increasing group and the decreasing group meet at $e\simeq0.5$, which means the crossings of the self-energy take place near at $e=0.5$. That is,
\begin{equation} \label{self-eng}
\Delta S_{e}:=|\omega_{jj}-\omega_{kk}|=0.
\end{equation}

\begin{figure}
\centering
\includegraphics[width=9.0cm]{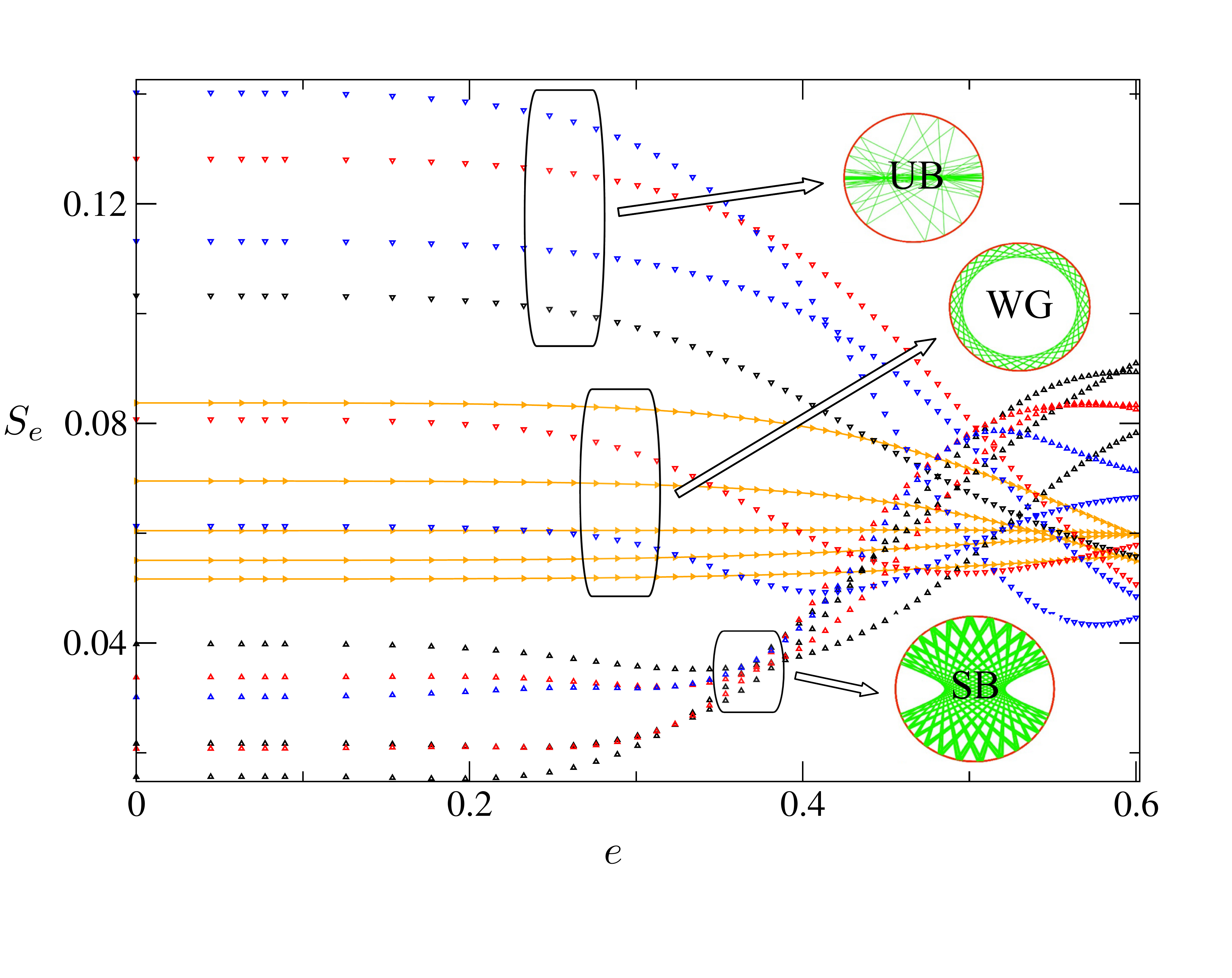}
\caption{(Color online) The transitions of self-energy $S_{e}$ of Fig.~\ref{Figure-3} depending on the eccentricity $e$. There are primarily three types of the self-energy transitions: whispering-gallery (WG, represented by $\triangleright$), stable-bouncing  (SB, represented by $\vartriangle$), and unstable-bouncing (UB, represented by $\triangledown$) modes, respectively. The SBs exhibit increasing pattern of self-energy, UBs exhibit decreasing of self-energy, and WGs exhibit almost unchanging of self-energy, respectively. In the case of bouncing-ball-type modes, the figures also show that the transition rate of the self-energy is rapidly increasing beyond $e=0.3$, and the crossings of the self-energy take place near  $e=0.5$.}
\label{Figure-4}
\end{figure}

\begin{figure}
\centering
\includegraphics[width=9.0cm]{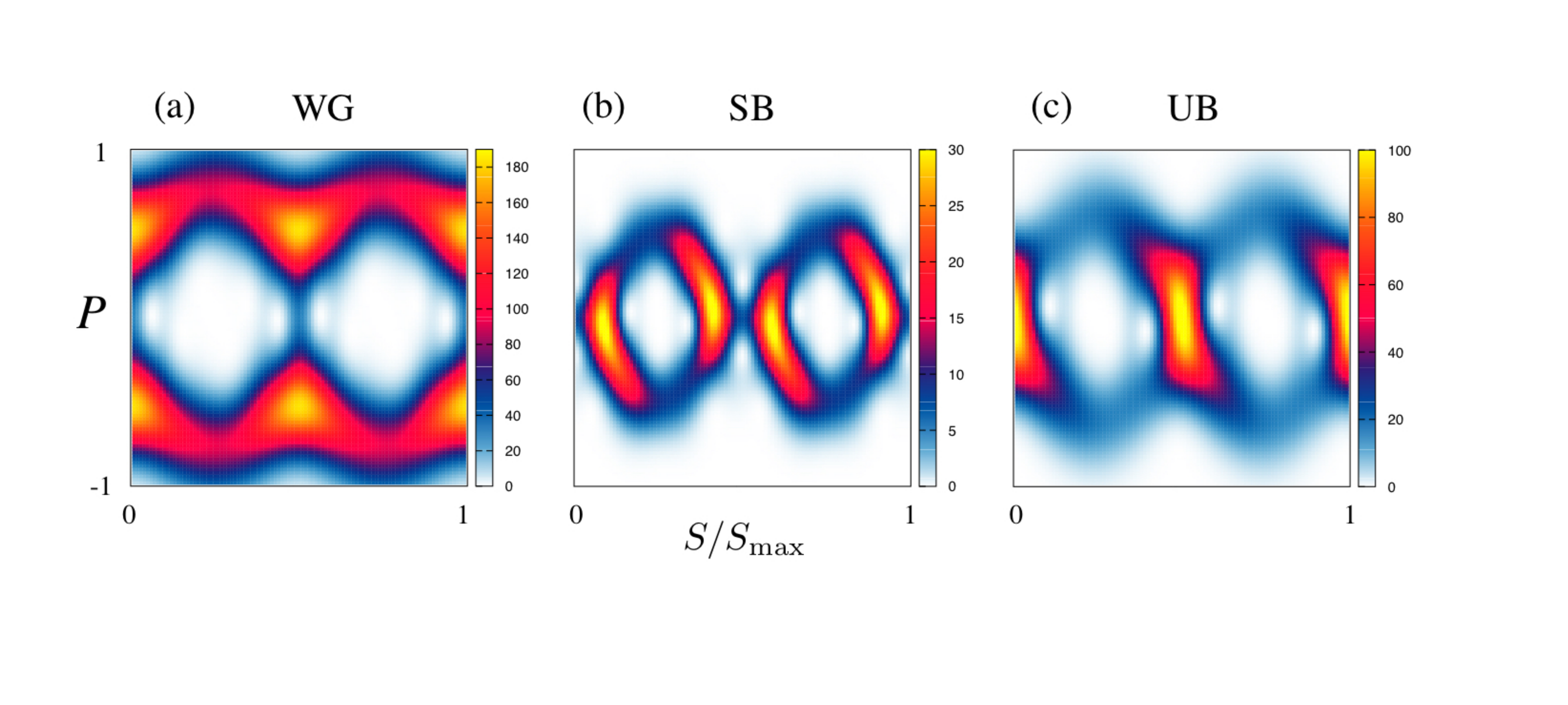}
\caption {(Color online) The Husimi probability distributions superimposed on PSOS. (a) One of the unchanging self-energy groups, i.e., $\ell=1,m=7$. This whispering-gallery mode is well localized on the intact invariant curves. (b) One of the increasing self-energy groups, i.e., $\ell=5,m=5$. This resonance mode is well localized on stable-bouncing-ball region. (c) One of the decreasing self-energy groups, i.e., $\ell=3,m=5$. This resonance mode is well localized on the unstable-bouncing-ball region.}
\label{Figure-5}
\end{figure}

First, we study how the three types of transitions can be classified. In order to do that, we get the Husimi probability distributions superimposed by PSOS on all of them at $e=0.6$.
As a result, these distributions are classified into three types of motions according to whispering-gallery (WG) modes, stable-bouncing (SB) ball modes, and unstable-bouncing (UB) modes. We plot one of the Husimi probability distributions in each group. Figure~\ref{Figure-5}(a) is one of the unchanging group, i.e., $(\ell=1,m=7)$. This resonance is well localized on the intact invariant curves. Figure~\ref{Figure-5}(b) is one of the increasing group, i.e., $(\ell=5,m=5)$.
This resonance mode is well localized on the stable-bouncing ball region. Figure~\ref{Figure-5}(c) is one of the decreasing group, i.e., $(\ell=3,m=5)$. This resonance mode is also well localized on the unstable-bouncing-ball region. Even though we do not exhibit all of the Husimi probability distributions, we can conclude that without loss of generality, unchanging groups are corresponding to the whispering-gallery modes group, the increasing groups are corresponding to the stable-bouncing-ball modes group, and the decreasing groups are corresponding to the unstable-bouncing-ball modes group. The tilted Husimi probability distributions of bouncing-ball-type modes are attributed to breaking of time-reversal symmetry, but the origin is not still fully understood~\cite{UWH08,AMH08,H09}.

\begin{figure}
\centering
\includegraphics[width=9.0cm]{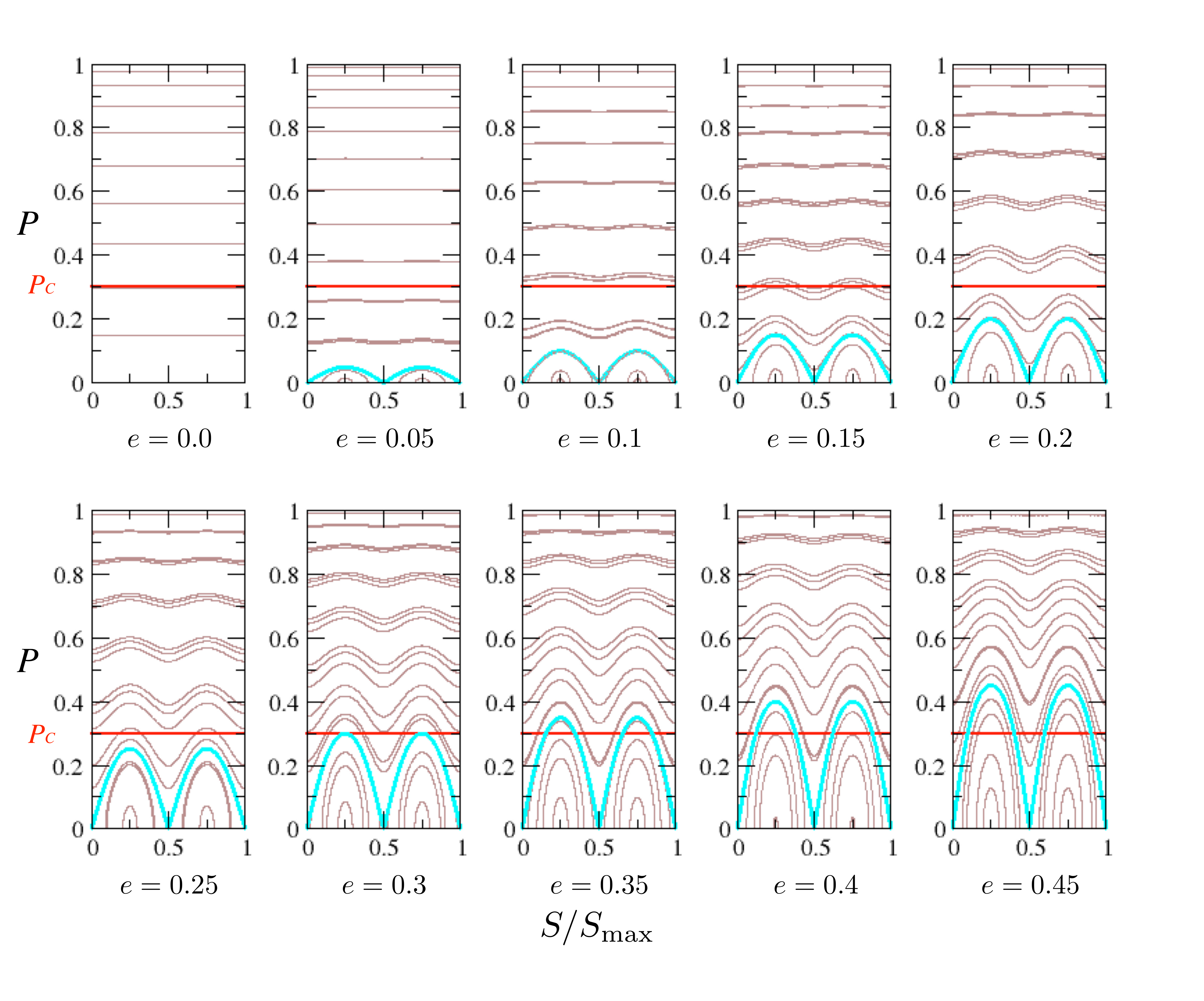}
\caption {(Color online) The Poincar\'{e} surface of sections with separatrixes and the critical lines at $e=0.0$, $e=0.05$, $e=0.1$, $e=0.15$, $e=0.2$, $e=0.25$, $e=0.3$, $e=0.35$, $e=0.4$, and $e=0.45$, respectively. The critical lines whose colors are red are placed near $P_{c}=0.3$. The stable-island structure is becoming larger as $e$ increases. Thus the separatrix whose colors are cyan are going up as $e$ increases. The separatrix touches the critical line near $e\simeq0.3$ and keeps soaring into the critical line beyond $e\simeq0.3$.}
\label{Figure-6}
\end{figure}

Second, we investigate why the transition rate of bouncing-ball-mode groups can be abruptly increased near $e=0.3$. It can be explained by PSOS with critical lines and separatrixes depending on the eccentricity $e$. In Fig.~\ref{Figure-6}, there are PSOSs of the elliptic billiard with separatrixes and critical lines at $e=0.0$, $e=0.05$, $e=0.1$, $e=0.15$, $e=0.2$, $e=0.25$, $e=0.3$, $e=0.35$, $e=0.4$, and $e=0.45$. The critical lines whose colors are red are placed near $P_{c}=0.3$. In the classical regime, i.e., ray dynamics, the light above this critical line cannot be leaked out. These separatrixes whose colors are cyan are border lines between the whispering-gallery motions and bouncing-ball motions. We also notice that the stable island structure is becoming larger as $e$ increases. Thus it means that the separatrixes are also going up as $e$ increases. These separatrixes still remain below the critical line until $e=0.3$ but they touch the critical line near $e=0.3$ and keep soaring into the critical line beyond $e=0.3$. This fact implies that the decay channels of the bouncing-ball-type modes undergo severe transitions during the separatrixes crossing the critical line. Because the self-energy is defined for each decay channel, these transitions of decay channels explain the transitions of self-energy. In the case of whispering-gallery motions, since the invariant curves in PSOS are nearly unchanged compared to bouncing-ball-type motions, their global transitions are also nearly unchanged compared to them.

%%%
\section{Self-energy versus Husimi distribution function} \label{Husimi}
The crossings of the self-energy $\Delta S_{e}=0$ for the bouncing-ball group take place near $e=0.5$. Our conjecture is that, since the self-energy is defined for each decay channel, the crossings of the self-energy may take place when a pair of compared resonances shares larger common decay channels.

\begin{figure}
\centering
\includegraphics[width=9.0cm]{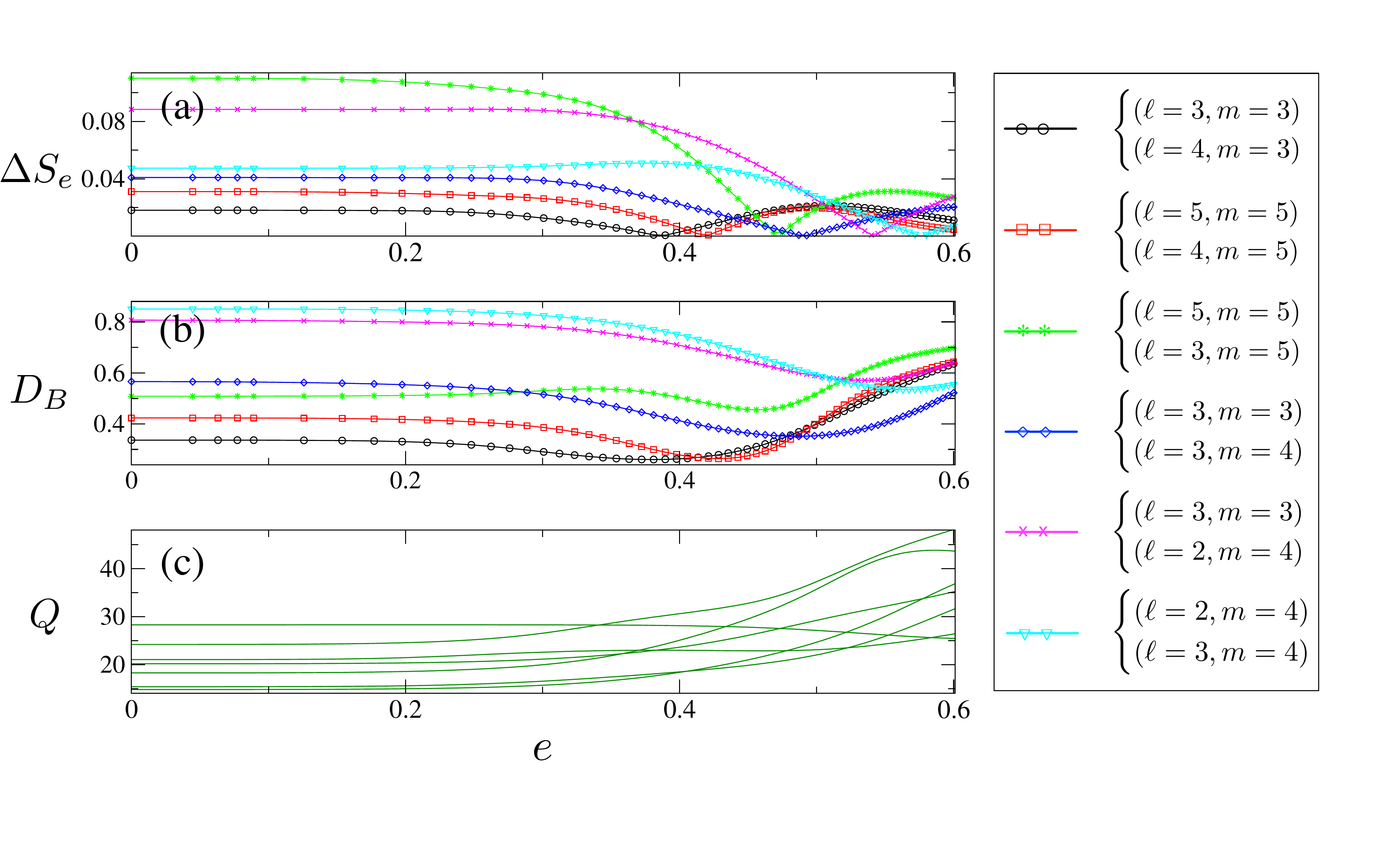}
\caption {(Color online) (a) Several selected relative differences of the self-energy ($\Delta S_{e}$) depending on $e$. The black curve ($\circ$) is one for $\Delta S_{e}$ between originating from $\ell=3,m=3$ and $\ell=4,m=3$. The red curve ($\Box$) is one for $\Delta S_{e}$ between originating from $\ell=5,m=5$ and $\ell=4,m=5$, and the green ($\ast$), blue ($\Diamond$), magenta ($\times$), and cyan ($\bigtriangledown$) curves are ones for $\Delta S_{e}$ between originating from $[(\ell=5,m=5),(\ell=3,m=5)]$, $[(\ell=3,m=4),(\ell=3,m=3)]$, $[(\ell=3,m=3),(\ell=2,m=4)]$, and $[(\ell=2,m=4),(\ell=3,m=4)]$, respectively. (b) The Bhattacharyya distance $(D_{B})$ of Husimi probability distributions below the critical line associated with same pair of resonances in $\Delta S_{e}$. This relation directly measures the similarity of decay channels. (c) The quality factors $Q$ associated with the same pair of resonances in $\Delta S_{e}$.}
\label{Figure-7}
\end{figure}

In order to confirm that, we select five pairs of relative differences of self-energy ($\Delta S_{e}=|\omega_{jj}-\omega_{kk}|$) depending on $e$. These curves are shown in Fig.~\ref{Figure-7}(a). The black curve ($\circ$) is for $\Delta S_{e}$ between originating from $\big[(\ell=3,m=3)$, ($\ell=4,m=3)\big]$. The red curve ($\Box$) is for $\Delta S_{e}$ between originating from $\big[(\ell=5,m=5),(\ell=4,m=5)\big]$, and the green ($\ast$), blue ($\Diamond$), magenta ($\times$), and cyan ($\bigtriangledown$) curves are for $\Delta S_e$ between originating from $\big[(\ell=5,m=5),(\ell=3,m=5)\big]$, $\big[(\ell=3,m=3),
(\ell=3,m=4)\big]$, $\big[(\ell=3,m=3),(\ell=2,m=4)\big]$, and
$\big[(\ell=2,m=4),(\ell=3,m=4)\big]$, respectively. Even though all these values
are almost unchanged until $e=0.3$, beyond this point, they decrease and have zero
values and then increase again. The zero point of $\Delta S_{e}$ ($\Delta S_{e}=0$) for black ($\circ$) is near $e=0.38$ and for red ($\Box$) is near $e=0.42$, for green ($\ast$) is $e=0.47$, for blue ($\Diamond$) is $e=0.49$, for magenta ($\times$) is $e=0.54$, and for cyan ($\bigtriangledown$) is $e=0.57$.

Since the decay channels are controlled by Husimi distributions below the critical line~\cite{SG+10,SGLX14,WH08}, we measure the similarity of Husimi distributions below the critical line to define the degree of sharing of a common decay channel. For this reason, we employ the Bhattacharyya distance~\cite{B43,MMPZ08} for investigating the similarity of the decay channels (Husimi distributions below the critical line). The Bhattacharyya distance measures the similarity of different two probabilities. For probability distribution $p(x)$ and $q(x)$, it is defined by
\begin{align}
D_{B}(p(x),q(x))=-\ln\big[\kappa_B\big(p(x),q(x)\big)\big],
\end{align}
where the factor $\kappa_B$ (namely, Bhattacharyya coefficient) is given by
\begin{align}
\kappa_B\big(p(x),q(x)\big)=\int\sqrt{p(x)q(x)}dx.
\end{align}
Note that the Bhattacharyya coefficient measures the amount of overlap between two statistical populations, i.e., classical fidelity~\cite{MMPZ08}.

\begin{figure}
\centering
\includegraphics[width=9.5cm]{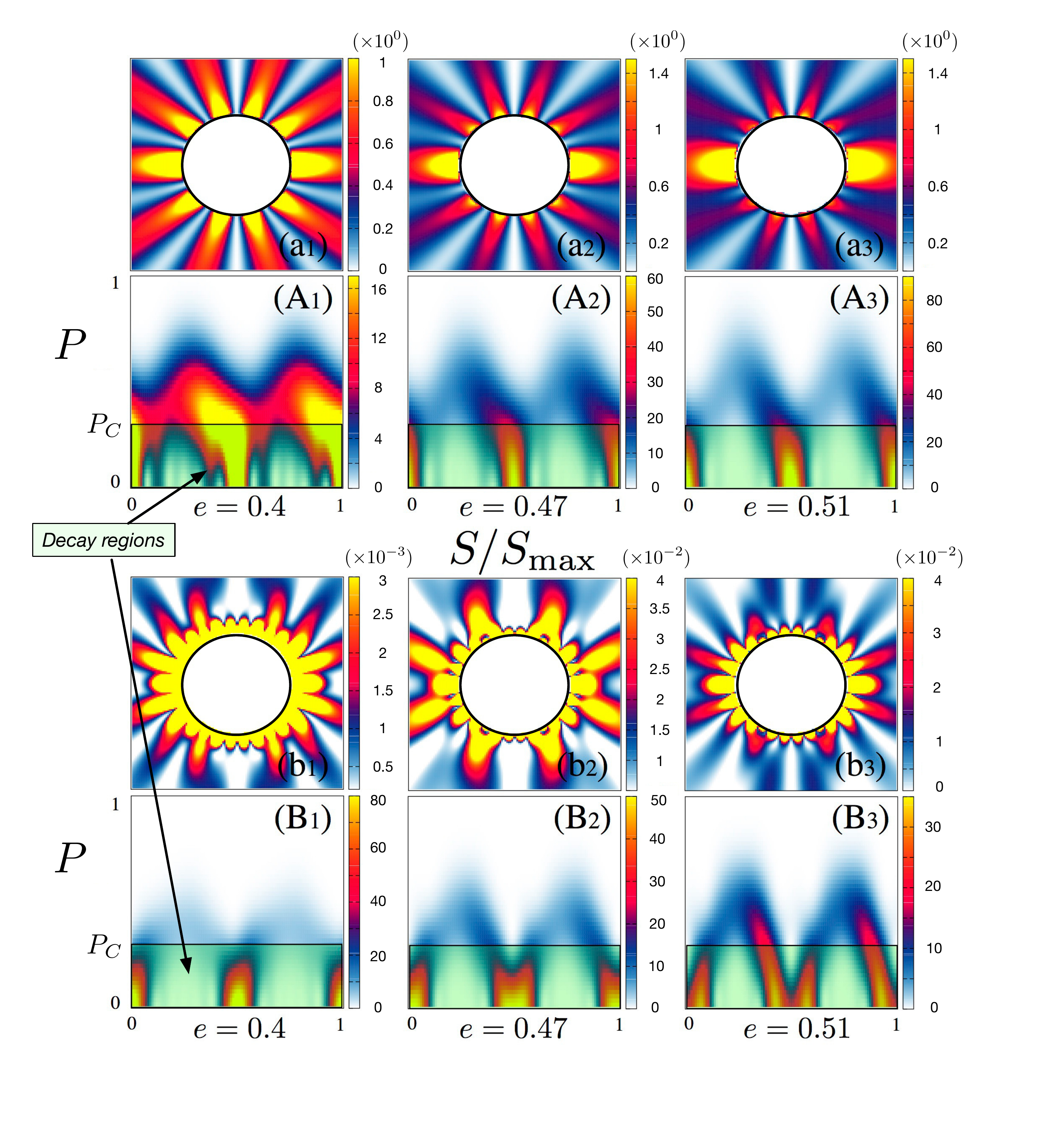}
\caption {(Color online) The intensity plots of resonance tails and Husimi probability distributions for $(\ell=3,m=5)$ in the indexes (a$_j$) and (A$_j$), and $(\ell=5,m=5)$ in (b$_j$) and (B$_j$) at $e=0.4$, $e=0.47$, and $e=0.51$, respectively. The Husimi probability distributions are very similar at $e=0.47$. This fact confirms that the Bhattacharyya distance $D_{B}$ has the lowest value at $e=0.47$.}
\label{Figure-8}
\end{figure}

In Fig.~\ref{Figure-7}(b), there are Bhattacharyya distances $D_{B}$ obtained from the decay channels associated with the same pair of resonances in $\Delta S_{e}$ of Fig.~\ref{Figure-7}(a). The global transition patterns are very similar to those of $\Delta S_{e}$. That is, the global transition patterns of $\Delta S_{e}$ and $D_{B}$ show a narrow parabola and broad parabola at the same region, respectively. Note that each extremal points of curves $\Delta S_{e}$ and $D_{B}$ are almost coincident at each $e=0.38$, $e=0.42$, $e=0.47$, $e=0.49$, $e=0.54$, and $e=0.57$. We also notice in Fig.~\ref{Figure-7}(c) that the quality factors of resonances involved in Fig.~\ref{Figure-7}(a) show the same order of magnitude.

There are intensity plots of eigenfunctions and Husimi probability distributions for $(\ell=5,m=3)$ with indexes (a$_j$) and (A$_j$) in Fig.~\ref{Figure-8}, and $(\ell=5,m=5)$ with (b$_j$) and (B$_j$) of Fig.~\ref{Figure-8} at $e=0.4$, $e=0.47$, and $e=0.51$, respectively. The plots (a$_j$) and (b$_j$) are resonance tails determined by Husimi probability distributions below the critical line represented by green shadow regions. These resonance tails play the role of decay channels~\cite{SGLX14,PKJ16}. We can easily notice that the green shadow regions are very similar at $e=0.47$. This fact confirms that the Bhattacharyya distance $D_{B}$ has the lowest value at $e=0.47$.

\begin{figure}
\centering
\includegraphics[width=9.0cm]{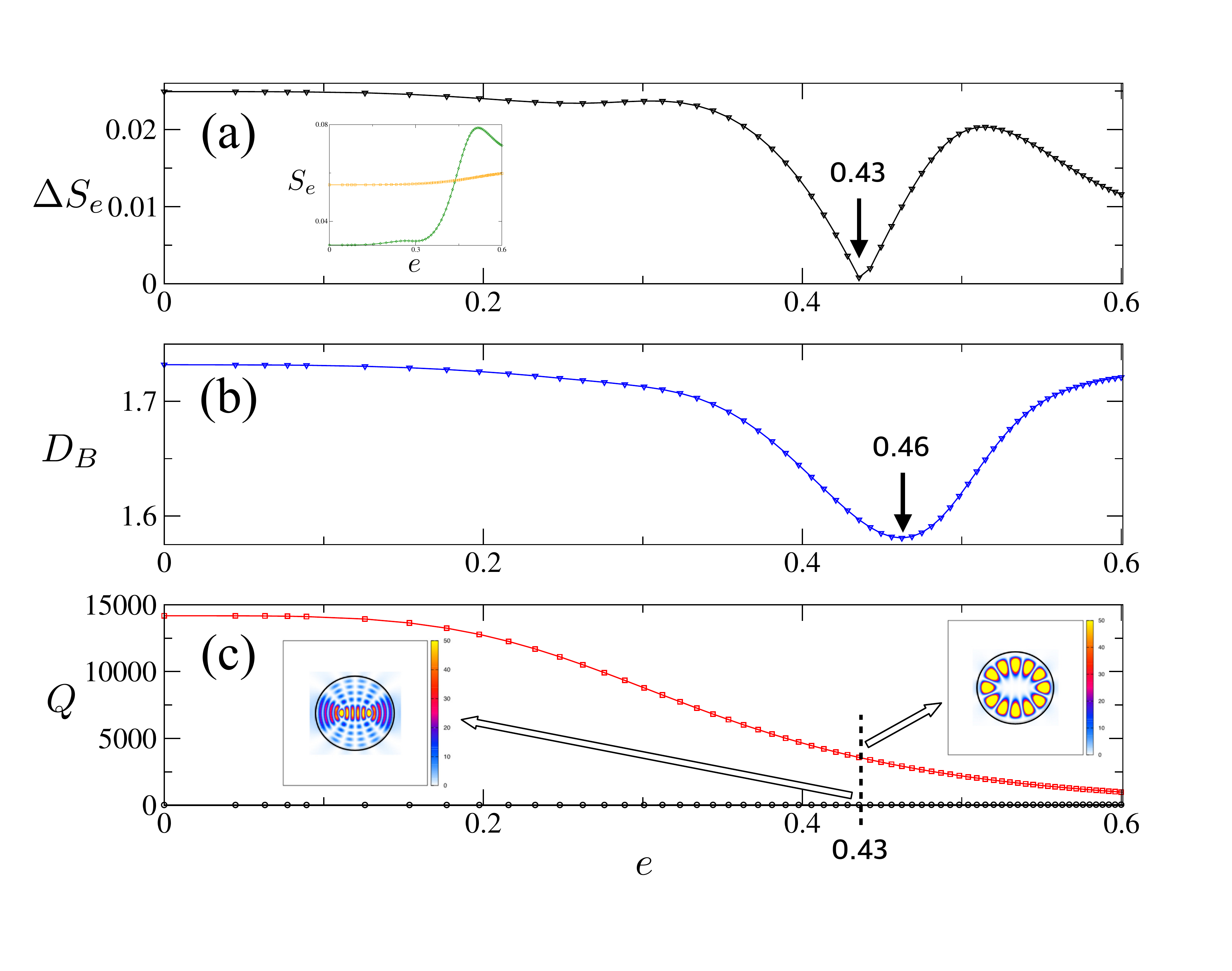}
\caption {(Color online) (a) Relative differences of self-energy ($\Delta S_{e}$) between single-layer WGMs ($\ell=1$ and $m=6$) and the stable-bouncing-ball-type mode ($\ell=5$ and $m=5$) depending on $e$. Its minimum value is lying at $e=0.43$. (b) The Bhattacharyya distance $D_{B}$ of the Husimi distribution below the critical line associated with same pair of resonances in $\Delta S_{e}$. Its minimum value is at $e=0.46$. The absolute magnitude of $D_{B}$ is much larger than those of Fig.~\ref{Figure-7}(b). (c) The quality factors ($Q$) associated with same pair of resonances in $\Delta S_{e}$. These two quality factors do not have similar order of magnitude.}
\label{Figure-9}
\end{figure}

We also investigate the relation of $\Delta S_{e}$ and $D_{B}$ between the single-layer WGM
($\ell=1$ and $m=6$) and stable-bouncing-ball-type mode ($\ell=5$ and $m=5$) depending on the eccentricity $e$. Their extremal positions are lying at $e\simeq0.43$ and $e\simeq0.46$, respectively. These are shown in Fig.~\ref{Figure-9}(a) and Fig.~\ref{Figure-9}(b). For panel (a) of Fig. 9, the linear (orange) curve is $S_{e}$ for $\ell=1,m=6$ and the S-shaped (green) one is $S_{e}$ for $\ell=5,m=5$. The quality factors $Q$ associated with the same pair of resonances in $\Delta S_{e}$ of Fig.~\ref{Figure-9}(a) are shown in Fig.~\ref{Figure-9}(c) and these two quality factors do not have same order of magnitude. We can figure out that even though the quality factors of the compared resonances are rather different from each other, if they share a common decay channel in \textit{narrow} parameter range, i.e., the parabola is manifested in the Bhattacharyya distance $D_{B}$ curve, the global transition of $\Delta S_{e}$ is strongly controlled by $D_{B}$. According to the results of Fig.~\ref{Figure-7} and Fig.~\ref{Figure-9}, we conjecture that if their quality factors have more similar orders of magnitude, the extremal points of $S_{e}$ and $D_{B}$ will be more coincident. Note that the absolute magnitude of $D_{B}$ between the WGM  and stable-bouncing-ball mode is much larger than those between bouncing-ball-type modes in Fig. 7(a). This is a natural result, because the similarity between the two bouncing-ball-type modes is much larger than that between the WGM and stable-bouncing-ball modes.

%%%
\section{Conclusion} \label{conclusion}
We have studied the geometrical boundary dependence of the Lamb shift (self-energy) through global transitions in an elliptic microcavity. This elliptic microcavity is a good platform to clearly see the Lamb shift, because elliptic billiards belong to an integrable system which has no internal interaction. We select the self-energy region by choosing without any avoided resonance crossings (ARCs) region. We confirm that these three types as a global transition correspond to the whispering-gallery modes, stable-bouncing-ball modes, and unstable-bouncing-ball modes by Husimi distributions superimposed by the classical Poincar\'{e} surface of sections. These facts directly indicate that the decay channel of an elliptic microcavity also can be classified into these three types of modes, because the self-energy is defined for each decay channel. The rate of global transition of self-energy is abruptly increased beyond $e\simeq0.3$, since the separatrix touches the critical line near at $e=0.3$, which means that the decay channels of bouncing-ball motions are abruptly changed around this region.

We have also investigated the crossings of self-energy, i.e., $\Delta S_{e}=0$. These self-energy crossings take place in the region where the resonances share much larger common decay channels in a \textit{narrow} parameter range. We confirm this fact by the Bhattacharyya distance $D_{B}$. The curves of relative differences of the self-energy ($\Delta S_{e}$) and the Bhattacharyya distance ($D_{B}$) show very similar patterns, and the extremal points of the parabola of these two curves are almost coincident. The extremal positions of the two curves $\Delta S_e$ and $D_B$ are getting closer as the difference between the quality factor of the two modes gets smaller. This fact implies that $\Delta S_e$ depends more on $D_B$ as the difference of the interaction of the system-environment of the two modes gets smaller. Our results give not only an understanding of the Lamb shift in the microcavity but also the implications of the openness effect theoretically and systemically.

%%%%
\section{acknowledgement}
We are grateful to Myung-Woon Kim, Sunghwan Rim, and Jung-Wan Ryu for comments. This work was partly supported by the IT R\&D program of MOTIE/KEIT (10043464). We thank the Korea Institute for Advanced Study for providing computing resources (KIAS Center for Advanced Computation Linux Cluster) for this work. K.J. acknowledges financial support by the National Research Foundation of Korea (NRF) through a grant funded by the Korea government (MSIP) (Grant No. 2010-0018295), by the KIST Institutional Program (Project No. 2E26680-16-P025), and the Associate Member Program funded by the Korea Institute for Advanced Study.


\begin{thebibliography}{26}

\bibitem{SHF11}
S. Shinohara, T. Harayama, and T. Fukushima,
%{Fresnel filtering of Gaussian beams in microcavities},
Opt. Lett.~\textbf{36}, 1023 (2011).

\bibitem{RTSCS02}
N. B. Rex, H. E. Tureci, H. G. L. Schwefel, R. K. Chang, and A. D. Stone,
%{Fresnel Filtering in Lasing Emission from Scarred Modes of Wave-Chaotic Optical Resonators},
\prl~\textbf{88}, 094102 (2002).

\bibitem{SH06}
H. Schomerus and M. Hentschel,
%{Correcting Ray Optics at Curved Dielectric Microresonator Interfaces: Phase-Space Unification of Fresnel Filtering and the Goos-H\"{a}nchen Shift},
\prl~\textbf{96}, 243903 (2006).

\bibitem{UWH08}
J. Unterhinninghofen, J. Wiersig, and M. Hentschel,
%{Goos-H\"{a}nchen shift and localization of optical modes in deformed microcavities},
\pre~\textbf{78}, 016201 (2008).

\bibitem{LRR+04}
S.-Y. Lee, S. Rim, J.-W. Ryu, T.-Y. Kwon, M. Choi, and C.-M. Kim,
%{Quasiscarred Resonances in a Spiral-Shaped Microcavity},
\prl~\textbf{93}, 164102 (2004).

\bibitem{LYM+09}
S.-B. Lee, J. Yang, S. Moon, S.-Y. Lee, J.-B. Shim, S. W. Kim, J.-H. Lee, and K. An,
%{Observation of an Exceptional Point in a Chaotic Optical Microcavity},
\prl~\textbf{103}, 134101 (2009).

\bibitem{H04}
W. D. Heiss,
%{Exceptional points of non-Hermitian operators},
J. Phys. A: Math. Gen.~\textbf{37}, 2455 (2004).

\bibitem{MR08}
M. M\"{u}ller and I. Rotter,
%{Exceptional points in open quantum systems},
J. Phys. A: Math. Gen.~\textbf{41}, 244018 (2008).

\bibitem{KKS99}
D. L. Kaufman, I. Kosztin, and K. Schulten,
%{Expansion method for stationary states of quantum billiards},
Am. J. Phys.~\textbf{67}, 133 (1999).

\bibitem{S99}
H.-J. St\"{o}ckmann,
\emph{Quantum Chaos: An Introduction}
(Cambridge University Press, London, 1999).

\bibitem{WWD97}
H. Waalkens, J. Wiersig, and H. R. Dullin,
%{Elliptic Quantum Billiard},
Ann. Phys.~\textbf{260}, 50 (1997).

\bibitem{LR47}
W. E. Lamb Jr., and R. C. Retherford,
%{Fine Structure of the Hydrogen Atom by a Microwave Method},
Phys. Rev. \textbf{72}, 241 (1947).

\bibitem{SS10}
M. O. Scully and A. A. Svidzinsky,
%{The Lamb Shift---Yesterday, Today and Tomorrow},
Science~\textbf{328}, 1239 (2010).

\bibitem{WKG04}
X.-H. Wang, Y. S. Kivshar, and B.-Y. Gu,
%{Giant Lamb Shift in Photonic Crystals},
\prl~\textbf{93}, 073901 (2004).

\bibitem{N10}
B. H. Nguyen,
%{Lamb and ac Stark shifts in cavity quantum electrodynamics},
Adv. Nat. Sci.: Nanosci. Nanotechnol.~\textbf{1}, 035008 (2010).

\bibitem{PKJ16}
K.-W. Park, J. Kim, and K. Jeong,
%{Non-Hermitian Hamiltonian and Lamb shift in circular dielectric microcavity},
Opt. Commun.~\textbf{368}, 190 (2016).

\bibitem{R09}
I. Rotter,
%{A non-Hermitian Hamilton operator and the physics of open quantum systems},
J. Phys. A: Math. Theor.~\textbf{42}, 153001 (2009).

\bibitem{D00}
F.-M. Dittes,
%{The decay of quantum systems with a small number of open channels},
Phys. Rep.~\textbf{339}, 215 (2000).

\bibitem{CK09}
G. L. Celardo and L. Kaplan,
%{Superradiance transition in one-dimensional nanostructures: An effective non-Hermitian Hamiltonian formalism},
\prb~\textbf{79}, 155108 (2009).

\bibitem{WKH08}
J. Wiersig, S. W. Kim, and M. Hentschel,
%{Asymmetric scattering and nonorthogonal mode patterns in optical microspirals},
\pra~\textbf{78}, 053809 (2008).

\bibitem{RWSS12}
R. R\"{o}hlsberger, H.-C. Wille, K. Schlage, and B. Sahoo,
%{Electromagnetically induced transparency with resonant nuclei in a cavity},
Nature~\textbf{482}, 199 (2012).

\bibitem{RSSCR10}
R. R\"{o}hlsberger, K. Schlage, B. Sahoo, S. Couet, and R. R\"{u}ffer,
%{Collective Lamb Shift in Single-Photon Superradiance},
Science~\textbf{328}, 1248 (2010).

\bibitem{R13}
I. Rotter,
%{Dynamical stabilization and time in open quantum systems},
Fortschr. Phys.~\textbf{61}, 178 (2013).

\bibitem{T89}
M. Tabor,
\emph{Chaos and integrability in nonlinear dynamics: An introduction}
(Wiley, New York, 1989).

\bibitem{N97}
J. U. N\"{o}ckel,
%{Resonances in Non Integrable Open Systems},
Yale University (Ph.D. Thesis), 1997.

\bibitem{M11}
N. Moiseyev,
\emph{Non-Hermitian Quantum Mechanics}
(Cambridge University Press, London, 2011).

\bibitem{RLK09}
J.-W. Ryu, S.-Y. Lee, and S. W. Kim,
%{Coupled nonidentical microdisks: Avoided crossing of energy levels and unidirectional far-field emission},
\pra~\textbf{79}, 053858 (2009).

\bibitem{J06}
J. Wiersig,
%{Formation of Long-Lived, Scarlike Modes near Avoided Resonance Crossings in Optical Microcavities},
\prl~\textbf{97}, 253901 (2006).

\bibitem{SGWC13}
Q. Song, L. Ge, J. Wiersig, and H. Cao,
%{Formation of long-lived resonances in hexagonal cavities by strong coupling of superscar modes},
\pra~\textbf{88}, 023834 (2013).

\bibitem{H01}
F. Haake,
\emph {Quantum Signatures of Chaos}
(Springer, Berlin, 2010).

\bibitem{AMH08}
E. G. Altmann, G. Del Magno, and M. Hentschel,
%{Non-Hamiltonian dynamics in optical microcavities resulting from wave-inspired corrections to geometric optics},
Europhys. Lett.~\textbf{84}, 10008 (2008).

\bibitem{H09}
M. Hentschel, 
%{Optical Microcavities as Quantum-Chaotic Model Systems: Openness Makes the Difference!}
Adv. Solid State Phys.~\textbf{48}, 293 (2009).

\bibitem{SG+10}
Q. H. Song, L. Ge, A. D. Stone, H. Cao, J. Wiersig, J.-B. Shim, J. Unterhinninghofen, W. Fang, and G. S. Solomon,
%{Directional Laser Emission from a Wavelength-Scale Chaotic Microcavity},
\prl~\textbf{105}, 103902 (2010).

\bibitem{SGLX14}
Q. Song, Z. Gu, S. Liu, and S. Xiao,
%{Coherent destruction of tunneling in chaotic microcavities via three-state anti-crossings},
Sci. Rep.~\textbf{4}, 4858 (2014).

\bibitem{WH08}
J. Wiersig and M. Hentschel,
%{Combining Directional Light Output and Ultralow Loss in Deformed Microdisks},
\prl~\textbf{100}, 033901 (2008).

\bibitem{B43}
 A. Bhattacharyya,
%{On a measure of divergence between two statistical populations defined by their probability distributions},
Bull. Cal. Math. Soc.~\textbf{35}, 99 (1943).

\bibitem{MMPZ08}
D. Markham, J. A. Miszczak, Z. Pucha{\l}a, and K. \.{Z}yczkowski,
%{Quantum state discrimination: A geometric approach},
\pra~\textbf{77}, 042111 (2008).


\end{thebibliography}
\end{document}